\documentclass[10pt, 5p, twocolumn]{elsarticle}

\usepackage[colorlinks=true,linkcolor=blue]{hyperref}
\usepackage{amsmath}
\usepackage{natbib}
\usepackage{graphicx}
\usepackage{dcolumn}
\usepackage{stix}
\usepackage{xcolor}
\definecolor{Fe100MO}{RGB}{0, 0, 0}
\definecolor{Fe100MAO}{RGB}{0, 0, 255}
\usepackage{bm}
\usepackage{stix}

\journal{Thin Solid Films}

\begin{document}
\begin{frontmatter}
\title{Influence of misfit strain on the physical properties of Fe thin films}

\author{Anna L. Ravensburg}
\author{Gunnar K. P\'alsson}
\author{Merlin Pohlit}
\author{Bj\"orgvin Hj\"orvarsson}
\author{Vassilios Kapaklis}

\address{Department of Physics and Astronomy, Uppsala University, Box 516, SE-75120 Uppsala, Sweden}

\begin{abstract}
We investigate the growth of thin Fe layers on MgAl$_2$O$_4$~(001) and MgO~(001) substrates using dc magnetron sputtering. The crystal quality of Fe layers deposited on MgAl$_2$O$_4$ is found to be substantially higher as compared to Fe grown on MgO substrates. The effects of the crystal quality on the magnetic and electronic transport properties are discussed.
\end{abstract}


\end{frontmatter}



\section{\label{sec:Introduction}Introduction}
 
\begin{figure}
\includegraphics{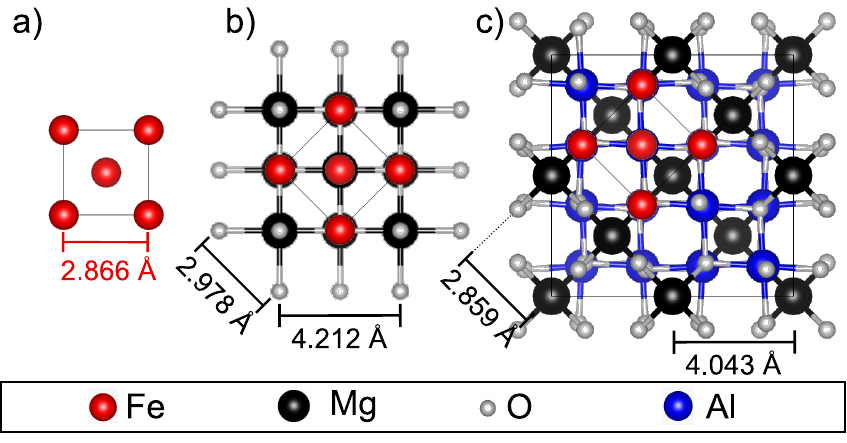}
\caption{\label{fig:motivation}Schematic illustration of a) a bcc Fe unit cell ($a_{\text{Fe}}$~=~2.866~{\AA} \cite{LandoltBornstein1994, Vassent1996}) and its orientation relative to the substrate if Fe is grown as a thin film b) on MgO~(001) ($a_{\text{MgO}}$~=~4.212~{\AA} \cite{LandoltBornstein1994}) or c) on MgAl$_2$O$_4$~(001) ($\frac{1}{2}$ $a_{\rm{MgAl_2O_4}}$~=~4.043~{\AA} \cite{LandoltBornstein2004}). The Fe unit cell is rotated by 45~degrees relative to the substrate, i.e., in its <110> direction. This distance is 2.978~{\AA} and 2.859~{\AA} for MgO and MgAl$_2$O$_4$, respectively. Crystal structures were taken from \textsc{Vesta} \cite{Momma2008}.}
\end{figure}

Understanding and limiting the relaxation of strain, originating from a lattice mismatch between substrate and film, is essential for the obtained structural quality and stability of heteroepitaxially grown layers \cite{Wedler2004}. Furthermore, a change in crystal quality can give rise to changes in physical properties of the films. In this context, Fe layers grown on MgO can serve as a model system \cite{Urano1988, Benedetti2011, Balogh2013}. Layers of Fe~(001) can be grown epitaxially on MgO~(001) substrates \cite{LandoltBornstein1994, Dibona2002}, hereafter referred to as MgO for simplicity, despite the difference in lattice parameters of Fe and MgO which are $a_{\rm{Fe}}$~=~2.866~{\AA} \cite{LandoltBornstein1994, Vassent1996} and $a_{\rm{MgO}}$~=~4.212~{\AA} \cite{LandoltBornstein1994}, respectively. The difference in the atomic distance is mostly accommodated for by a rotation of the Fe unit cell with respect to MgO. The Fe~<100> and MgO~<110> axes become parallel, providing a growth condition with $\sim$+4~\% lattice mismatch \cite{Vassent1996, Meyerheim2001, Meyerheim2002} (see Fig. \ref{fig:motivation}a and b). The mismatch still implies significant tensile in-plane strain in Fe \cite{Muehge1994, Dibona2002, Vassent1996, Raanaei2008}, resulting in defect generation above the critical thickness and thereby potentially altering the physical properties of the layers. 
 
 Fe can be grown on MgAl$_2$O$_4$~(001) ($a_{\rm{MgAl_2O_4}}$~=~8.086~{\AA} \cite{LandoltBornstein2004}), hereafter referred to as MAO for simplicity, providing better lattice matching (-0.2~\%) \cite{Hosseini2008, Ganesh2013, Sukegawa2010} as compared to MgO ($\sim$+4~\%), see Fig. \ref{fig:motivation}c. The growth of thin Fe layers on MAO has been shown to result in a better crystal quality of the Fe layers, as compared to when grown on MgO \cite{Khodadadi2020}. This has been used for investigations of the effects of crystal quality on Gilbert damping \cite{Khodadadi2020}, THz emission as well as on the magnetic properties of Fe in Fe/MgAl$_2$O$_4$/Fe~(001) \cite{Sukegawa2010, Miura2012, Belmoubarik2016, Masuda2017, Xiang2018}.
Crystalline Fe has a fourfold in-plane magnetocrystalline anisotropy. The Fe easy axes are along the Fe~<001> direction, while the Fe hard axes point along the Fe~<110> direction. 
The strain of epitaxially grown ferromagnets has been found to alter the magnetic anisotropy \cite{Chunhui2013, Li1999}, and electronic transport has been shown to be altered as well, for thicknesses larger than 110~{\AA} \cite{Khodadadi2020}. However, a systematic comparison of the growth, crystal quality, and the resulting physical properties of Fe are not available in the literature. 
 
Here we investigate how the choice of substrate and thickness of the Fe layers affects some of the physical properties. The thickness of the investigated layers is in the range 12.5 to 100~{\AA} and we address how the crystal quality as well as the magnetic and electronic transport properties are affected when Fe is grown on MAO and MgO substrates.

\section{\label{sec:ExperimentalDetails}Experimental details}

\subsection{\label{sec:Growth}Growth}

Fe thin films were simultaneously deposited on single crystalline MAO and MgO substrates (10$\times$10~mm$^2$), using dc magnetron sputtering, eliminating this way all uncertainties concerning differences in growth conditions. Prior to the growth, the substrates were annealed for 600~s at 1273(2)~K (, where the number in brackets shows the estimated uncertainty of the last digit). The target-to-substrate distance in the deposition chamber is approximately 0.2~m. The deposition rates (Fe: 0.1~{\AA}/s, Pd: 0.6~{\AA}/s) were calibrated prior to the growth of the samples. As seen in Table \ref{tab:CrystalStructure}, the actual Fe layer thicknesses lie reasonably close to the nominal values. Hence, for simplicity we hereafter refer to the samples by their intended nominal thicknesses. Fe layers with nominal thicknesses $t_{\text{Fe}}$ of 100~{\AA}, 50~{\AA}, 25~{\AA}, and 12.5~{\AA} were grown at a substrate temperature of 619(2)~K. The growth temperature was selected based on the analysis of a series of 100~{\AA} thick Fe layers at different deposition temperatures between 606(2) and 656(2)~K. Well-defined layering in combination with the highest crystal quality, lowest magnetic saturation field, and lowest coercive field have been observed for Fe grown at 619(2)~K independent of substrate. In order to prevent oxidation upon exposure to air, the films were capped by a 50~{\AA} thick Pd layer. The deposition of the capping layers was made at ambient temperatures.
The base pressure of the growth chamber is below 5$\times$10$^{-7}$~Pa. Both layers were sputtered at a power of 50~W in an Ar atmosphere (gas purity $\geq$~99.999~\%, and a secondary getter based purification) from targets with a diameter of 5.08~cm (purity: 99.95~\%) at a deposition pressure of 0.67 and 1.07~Pa for Fe and Pd, respectively. The targets were cleaned by sputtering with closed shutters for 60~s prior to each deposition. The substrate holder was rotated at 30~rpm during the deposition of the samples in order to ensure thickness uniformity. 

\subsection{\label{sec:Characterisation}Characterisation}

X-ray reflectometry (XRR) and diffraction (XRD) measurements were carried out using a Bede D1 equipped with a Cu x-ray source operated at 35~mA and 50~kV, a G\"obel mirror, and a 2-bounce-crystal on the incidence side. Furthermore, a circular mask (5~mm) as well as an incidence and a detector slit (both 0.5~mm) were used. The x-rays were detected with a Bede EDRc x-ray detector. The measured XRR data was fitted using \textsc{GenX} \cite{Bjorck2007} allowing for the determination of the thickness, roughness, and the scattering length density (SLD) profile of the layers. However, twinning in both, MAO and MgO substrates, and hence also in the epitaxially growing Fe layer, may lead to an overestimation of the layer roughnesses especially for thin Fe layers. XRD measurements around the substrate Bragg peaks confirmed substrate twinning for all samples except for the ones with 100~{\AA} Fe layer thickness on MAO and MgO. For the XRD studies, the samples were measured with coupled $2\theta$-$\theta$ scans and rocking curves around the Fe~(002) peak. In order to determine the full-width-at-half-maximum (FWHM) of the rocking curve peaks, the data was fitted with a Lorentzian. The instrument resolution of 0.012(1)~degrees was taken into account in the analysis. The critical Fe layer thicknesses of MAO/Fe and MgO/Fe films as well as the Fe Poisson's ratio were calculated. The calculations were performed based on the theory and formulas presented by Droulias and co-workers \cite{Droulias2017}. The considered elastic constants of Fe and the substrate materials were taken from elsewhere \cite{Droulias2017, Adams2006, Yoneda1990}.

Magnetization measurements were performed using a magneto-optical-Kerr-effect setup in longitudinal geometry with $s$-polarized light. The magnetic response is measured parallel to an in-plane applied magnetic field. The samples were probed along the <100> and <110> axes of Fe.

The temperature dependence of the resistivity in the films was determined using four-point probe measurements. The films were contacted by pressing four gold plated spring contacts in the collinear configuration on the sample surface with an equal spacing of 2.0(2)~mm between neighboring contacts. The measurements were performed using a standard dc current reversal technique, i.e., by sourcing a current $I$ of $\pm$ 100~$\mu$A with a Keithley 2400 sourcemeter and measuring the voltage $V$ using a Keithley 2182A nanovoltmeter. This selected method not only eliminates contact resistance but also thermally induced voltages. All measurements were performed during warm-up of the samples with a rate of 0.5~K/min, in order to reduce thermal gradients between sample and temperature sensor. Based on these measurements, the sheet resistance $R_{\square}$ of the Fe/Pd bilayer thin film was calculated using,
\begin{equation}
\label{equ:trans}
R_{\square}=\frac{\rho}{t}=\frac{\pi}{\ln{(2)}}\cdot \frac{V}{I}\cdot F
\end{equation}
with $F$~=~0.7744 \cite{Topsoe1968}. Note, that in the following the thickness $t$, resistivity $\rho$, and sheet resistance $R_{\square}$ are representing the Fe film together with the Pd capping layer. The two main sources of measurement error lie in the non-ideal film geometry, due to small clamp marks on the sample corners from the sputter growth and variations in the contact spacing when mounting a sample. The accuracy of the four-point probe method was validated on selected samples by using a van-der-Pauw configuration, with spring contacts brought as close as possible to the corners of the sample.  


\section{\label{sec:StructuralAnalysis}Results and Discussion}

\subsection{\label{sec:LayeringAndFilmThickness} Crystal quality and layering}

\begin{figure*}[t]
\includegraphics [scale = 1] {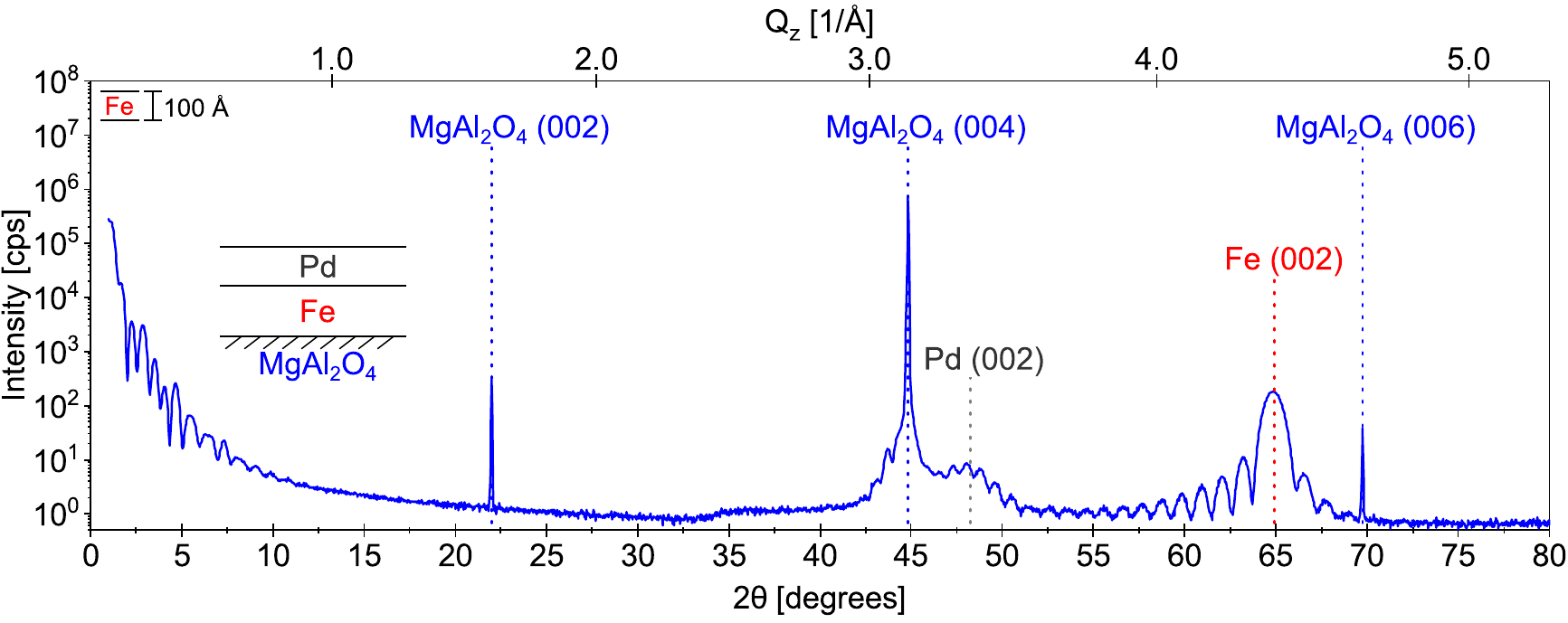}
\caption{\label{fig:Xrays} X-ray diffraction pattern, measured using CuK$_{\mathrm{\alpha}}$ radiation, of a Fe thin film deposited on MAO with a Fe layer thickness of 100~{\AA} and a 50~{\AA} Pd capping layer. The upper scale shows the momentum transfer ($Q = 4\pi \sin{\theta}/ \lambda$ ) for convenience.}
\end{figure*}

Fig. \ref{fig:Xrays} illustrates results from a $2\theta$-$\theta$ scan obtained from a 100~{\AA} Fe layer grown on MAO, allowing us to define contributions from all the involved length scales. Total film thickness oscillations, Kiessig fringes \cite{Kiessig1931}, are observed at small angles. The peaks at 22, 45, and 69~degrees originate from the single crystalline substrate, MAO. The broad peak at 48~degrees is attributed to the Pd capping layer. The Fe~(002) Bragg peak is observed at 65~degrees, with Laue oscillations visible on both sides. Laue oscillations and Kiessig fringes represent different length scales in these samples: The Kiessig fringes originate from both, the Fe and the Pd layer, while the Laue oscillations stem solely from the single crystalline Fe layer.
The observed Laue oscillations are asymmetric and their intensity distribution falls more abruptly on the high-angle side of the Fe~(002) peak. On the low-angle side, they are visible for a range of more than 20~degrees. Laue oscillations around the Fe~(002) peak are observed if the coherency of the structure is of the order of the total film thickness and hence, their occurrence reflects the high crystal quality of Fe grown on MAO \cite{Lee2017}.  

\begin{figure}
\includegraphics{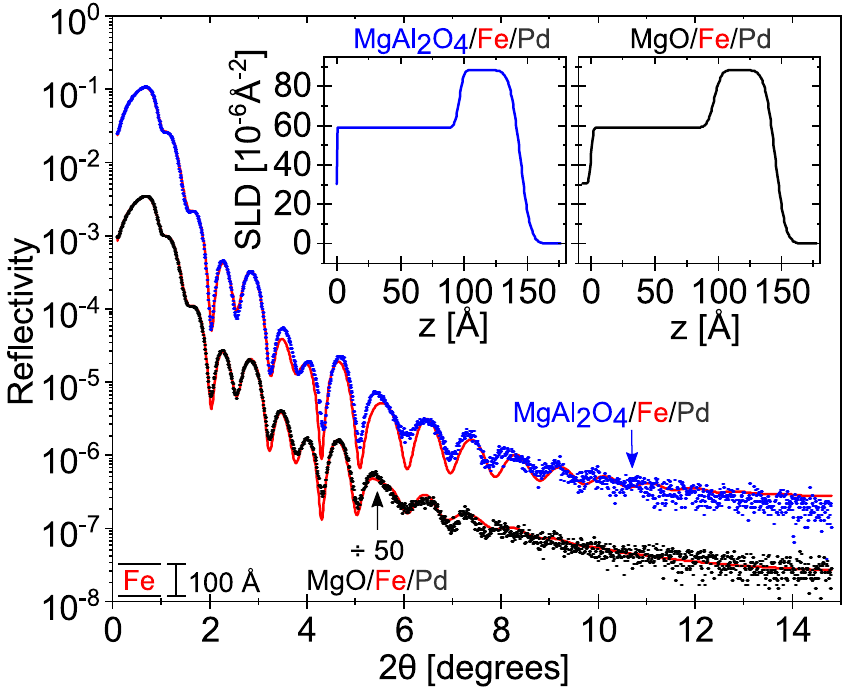}
\caption{\label{fig:Layering}X-ray reflectivity scans of thin films (100~{\AA} Fe) deposited on single crystalline MAO and MgO substrates. The Fe layers were capped with Pd at room temperature. The solid lines are fits obtained using \textsc{GenX} \cite{Bjorck2007} (red). The scan obtained from MgO/Fe/Pd is shifted for clarity by a factor of 50. The obtained SLD profiles of the samples are shown as insets.}
\end{figure}

We start the discussion on the structural quality by having a closer look on the layering of the films. Representative XRR scans for 100~{\AA} Fe grown on MAO and MgO are shown in Fig. \ref{fig:Layering}. The data was fitted using \textsc{GenX} \cite{Bjorck2007}. Kiessig fringes are visible up to 13~degrees for Fe on MAO and 9~degrees for Fe grown on MgO. The Kiessig fringes are more pronounced for Fe on MAO as compared to MgO, in line with the results on the fitted width of the Fe/Pd interfaces, see table \ref{tab:CrystalStructure}. The increased fitted roughness for the thinnest Fe layers on both substrates is attributed to the relatively larger contribution of the substrate/Fe interface. The single crystalline substrates may be twinned \cite{Schroeder2015} or have atomic terraces with incommensurate out-of-plane height in relation to the Fe atomic distance \cite{LandoltBornstein1994}. The SLD profiles obtained from the fitting of the samples with 100~{\AA} Fe layer thickness confirm the intended substrate/Fe/Pd layering with well defined layer thicknesses on both substrates. The profiles show a small difference in substrate and Fe/Pd interface roughness. It is smaller for the MAO compared to the MgO sample which is in agreement with the more pronounced Kiessig fringes for that sample.
The thickness of the Fe layers $t_{\text{Fe}}$ for the samples
were determined to be 96(1)~{\AA} and 97(1)~{\AA} for MAO and MgO, respectively. The intended and actual thickness of the layers are therefore within 5~\%.

\begin{table}
\caption{\label{tab:CrystalStructure}Results from fitting x-ray reflectivity with GenX \cite{Bjorck2007} of Fe thin films grown on MAO (a) to d)) and MgO (e) to h)). The Fe layer thickness ($t_{\text{Fe}}$) was varied while the thickness of the Pd capping layer ($t_{\text{Pd}}$) was kept constant. Additionally, substrate and layer roughnesses ($\sigma_{\text{Sub}}$, $\sigma_{\text{Fe}}$, and $\sigma_{\text{Pd}}$) were obtained from the fitting.}
\begin{tabular}{ |>{\centering\arraybackslash}m{0.3cm}|>{\centering\arraybackslash}m{1.2cm}|>{\centering\arraybackslash}m{1.15cm}|>{\centering\arraybackslash}m{1.15cm}|>{\centering\arraybackslash}m{1.15cm}|>{\centering\arraybackslash}m{1.15cm}|}
\hline
& \multicolumn{1}{c|}{Substrate} & \multicolumn{2}{c|}{Fe} & \multicolumn{2}{c|}{Pd} \\
\hline
& $\sigma_{\text{Sub}}$ [{\AA}] & $t_{\text{Fe}}$ [{\AA}] & $\sigma_{\text{Fe}}$ [{\AA}] & $t_{\text{Pd}}$ [{\AA}] & $\sigma_{\text{Pd}}$ [{\AA}] \\
\hline
\multicolumn{6}{|c|}{\textcolor{Fe100MAO}  {$\mdlgblkcircle$} Fe on MAO}\\
\hline
a) & 0(1) & 96(1) & 3(1) & 48(1) & 6(1)\\ 
b) & 0(1) & 48(1) & 1(1) & 46(1) & 4(1)\\
c) & 0(2) & 25(1) & 1(1) & 46(1) & 2(1)\\ 
d) & 0(2) & 11(1) & 7(2) & 48(1) & 1(1)\\ 
\hline
\multicolumn{6}{|c|}{\textcolor{Fe100MO}  {$\mdblksquare$} Fe on MgO}\\
\hline
e) & 1(1) & 97(1) & 4(1) & 48(1) & 7(1)\\
f) & 0(1) & 50(1) & 5(1) & 46(1) & 4(1)\\
g) & 0(2) & 26(1) & 2(1) & 46(1) & 3(1)\\
h) & 0(1) & 13(1) & 6(1) & 46(1) & 3(1)\\
\hline
\end{tabular}
\end{table}

\begin{figure}
\includegraphics{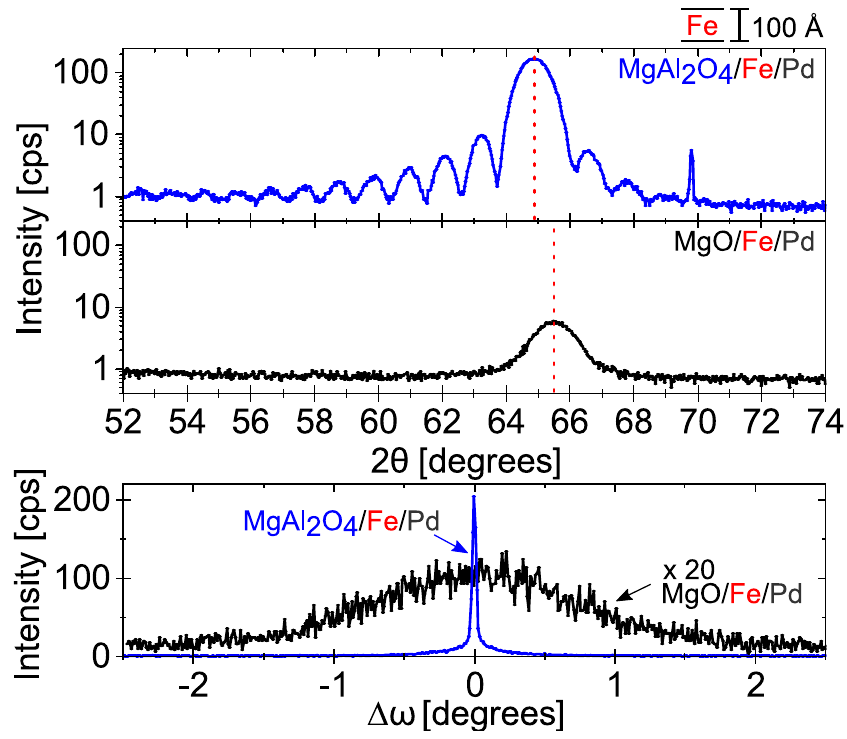}
\caption{\label{fig:CrystalStructure}$2\theta$-$\theta$ (top) and rocking curve (bottom) x-ray diffraction scans from Fe thin films deposited on MAO and MgO. The intensity of the rocking curve scan of MgO/Fe was multiplied by a factor of 20 for clarity.}
\end{figure}

Having established only minor impact of the substrate on the thickness variations of the Fe layers, we now have a look on how the choice of substrate affects the obtained crystal quality. A representative XRD scan of an Fe thin film with 100~{\AA} Fe layer thickness grown on MAO is shown in the upper plot in Fig. \ref{fig:CrystalStructure}. The sharp peak at 69~degrees originates from the MAO substrate. The Fe~(002) peak is clearly observed at 64.890(4)~degrees corresponding to an average out-of-plane lattice parameter of 2.8738(2)~{\AA}. Reciprocal space mapping around the Fe~(112) off-specular peak (not shown) allows for the determination of an in-plane lattice parameter being 2.8595(8)~{\AA}. Consequently, the in-plane parameter is smaller than the out-of-plane lattice parameter, giving rise to a tetragonal distortion of the cubic lattice. The measured Fe in-plane lattice parameter lies close to the corresponding in-plane atomic distance in the MAO substrate of 2.859~{\AA}, consistent with a fully strained growth for 100~{\AA} Fe on MAO. The elastic response of the 100~{\AA} Fe grown on MAO was calculated using the measured in- and out-of-plane lattice parameters as well as the Fe equilibrium lattice parameter of 2.866~{\AA} \cite{LandoltBornstein1994, Vassent1996}. The resulting Poisson's ratio of 0.37(4) is in good agreement with $\nu$~=~0.368 which was calculated using the elastic constants of bulk Fe.

An XRD scan of a 100~{\AA} Fe film grown on MgO is shown in the second plot of Fig. \ref{fig:CrystalStructure}. The out-of-plane lattice parameter was determined to be 2.8500(2)~{\AA}, which is smaller than obtained from Fe on MAO and smaller than the Fe equilibrium lattice parameter of 2.866~{\AA} \cite{LandoltBornstein1994, Vassent1996}. The lattice mismatch of Fe and MgO is $\sim$4~\% \cite{Vassent1996, Meyerheim2001, Meyerheim2002}, inducing tensile in plane strain resulting in a compressive out-of-plane Poisson response. 
The intensity of the Bragg peak is almost two orders of magnitude smaller for Fe on MgO, as compared to Fe on MAO. Corresponding difference is seen in the rocking curves of the Fe~(002) peaks displayed in the bottom plot of Fig. \ref{fig:CrystalStructure}. The width of the rocking curve, obtained from Fe grown on MAO, is one order of magnitude smaller compared to the Fe layers grown on MgO. The FWHM, and therefore the mosaicity, 
of the Fe~(002) rocking curve on MAO is below 0.05~degrees for all Fe layer thicknesses within this study. No trends in the width of the rocking curve are observed with layer thickness. The resolution of the measurements is 0.012(1)~degrees. 
A mosaic spread below 0.05~degrees is compelling evidence of the high crystal quality of Fe grown on MAO \cite{Muehge1994}, even for Fe layers of 12.5~{\AA} in thickness.
In contrast, a 100~{\AA} Fe layer on MgO exhibits an Fe~(002) rocking curve peak with a FWHM of 1.76~degrees, corresponding to a more than 40 times larger mosaic spread and significantly lower crystal quality compared to the layer on MAO. For thinner Fe layers on MgO, the mosaic spread increases significantly (up to 60~\%), becoming 2.84~degrees for 50~{\AA} thick Fe layers. Intensities of the peaks were below the detection limit of the instrument when the thickness of the Fe layers were below 50~{\AA}, if grown on MgO. Hence, large differences in crystal quality are obtained when Fe is grown on MAO and MgO, in stark contrast to the similar quality of the layering of Fe on the same substrates. 

Besides the intensity, the most significant difference in the diffraction patterns in Fig. \ref{fig:CrystalStructure} is the presence and absence of Laue oscillations. Laue oscillations can be observed when the coherence length of the sample is of the order of the film thickness and the latter is smaller than the spatial coherence length of the x-ray beam. The presence of Laue oscillations around the (002) peak of Fe provides therefore conclusive evidence for high degree of structural coherency as well as well defined thickness of the Fe layers grown on MAO \cite{Ying2009}. For 50~{\AA} as well as 25~{\AA} Fe on MAO, Laue oscillations are observed as well. For the thinnest sample in this study, 12.5~{\AA} Fe, the intensity was too low to allow detection of the Laue oscillations. The oscillations are asymmetric in intensity around the Bragg peak, decaying faster on the high angle side. This asymmetry can have multiple origins, one being a strain profile within the layer \cite{Vartanyants2000, Robinson2001} but also terraces in the single crystalline substrate and the deposited film, can cause such an asymmetry. 

No Laue oscillations are observed from Fe films grown on MgO for any sample within this study, which is in line with the results obtained by M\"uhge and co-workers \cite{Muehge1994}. Significant plastic strain relaxation is caused by misfit dislocations in the films \cite{Fitzgerald1991}, affecting the coherence length of the crystal \cite{Ying2009}. The formation of dislocations is expected above the critical thickness and screw dislocation formation in bcc Fe~(001) grown on MgO, for example, is experimentally confirmed for layer thicknesses above 25~{\AA} \cite{Wedler2004}. Consequently, profound strain is expected in Fe on MgO.
The calculated critical thickness $h_{c}$ of Fe on MAO is 785~{\AA} and, thus, significantly larger than the thickest Fe layer investigated here. Hence, Fe is expected to grow coherently (fully strained) on MAO since the lattice mismatch between bulk Fe and MAO is only -0.2~\% \cite{Sukegawa2010}. Consequently, the mosaicity is expected to be independent of the thickness of the Fe layer, which is consistent with our findings. In contrast, the critical thickness of Fe grown on MgO is calculated to be 20~{\AA}. Therefore, relaxation in Fe grown on MgO starts at significantly thinner Fe layers: >~20~{\AA}, misfit dislocation formation is expected for Fe grown on MgO. This should cause an increase in the mosaic spread with decreasing Fe layer thickness where the portion of the film with dislocations is larger. 
 For Fe layer thicknesses between 100~{\AA} and 300~{\AA} grown on MgO, the mosaic spread is reported to be rather constant \cite{Muehge1994} due to the large degree of relaxation for the initial 100~{\AA}.

\subsection{\label{sec:level2}Magnetic properties}

\begin{figure}
\includegraphics{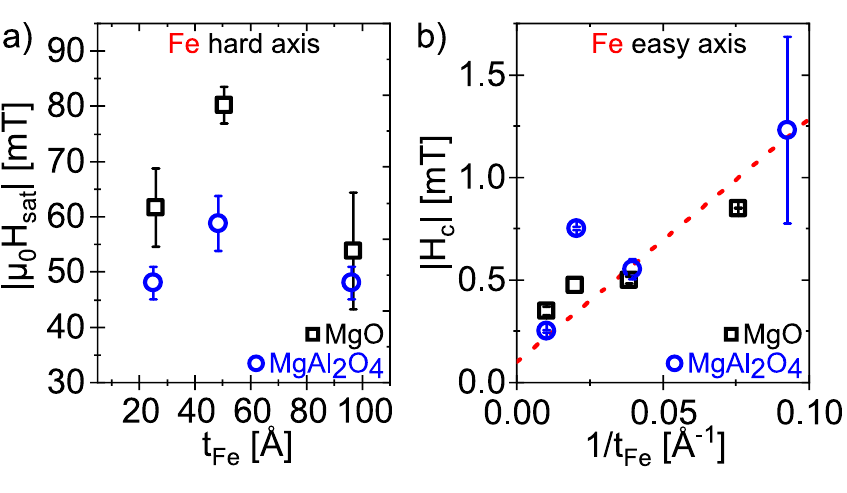}
\caption{\label{fig:Magnetism}a) Saturation field $|\mu_0H_{sat}|$ and b) absolute value of coercivity $|H_{c}|$ of Fe thin films deposited on MAO (blue) and MgO (black) while varying the Fe layer thickness. The measurements for a) were conducted with a magnetic field applied along the Fe hard axis while for the measurements in b) a magnetic field was applied along the Fe easy axis. The dashed straight line is a fit to the data while constraining the intersect to the field axis, at the coercivity value of bulk iron.}
\end{figure}

The saturation field along the hard in-plane axis of Fe, grown on MAO and MgO, is displayed in Fig. \ref{fig:Magnetism}a. Fe deposited on MAO appears to have smaller saturation field than Fe on MgO. Both are somewhat higher than the literature value of bulk Fe, which is approximately 45~mT \cite{Kittel1976}. The strain state in thin films is known to influence spin-orbit coupling and thereby the magnetic anisotropy \cite{Huang2016, Salikhov2017, GutierrezPerez2018}.
Hence, the obtained results are consistent with the observed difference in strain state in Fe on MAO and MgO, while the influence of thickness of the Fe layer on the saturation field appears as weak or non-conclusive. 

The coercive field in the direction of easy axis, as a function of the inverse Fe layer thickness, is shown in Fig. \ref{fig:Magnetism}b. The observed scaling with inverse thickness can be understood as arising from constant contribution from an interface, acting as pinning centre. The contribution from the interior of the film can be inferred from the y-intercept, being set to 0.1~mT, which is comparable to single crystal Fe \cite{Muller2003}. The interface between Fe and Pd is expected to be irrelevant in the context, due to the large magnetic susceptibility of Pd combined with negligible coercivity of the induced moment \cite{Hase2014}. Furthermore, a linear dependence on inverse layer thickness has been observed for Fe layers in the thickness range 130-480~{\AA} \cite{Bensehil2017}, further supporting the role of the substrate/Fe interface, where Fe is reported to bond to the oxygen of the MAO and MgO substrates \cite{Meyerheim2001, Miura2012, Masuda2017}. We therefore ascribe the dominating contribution to the coercivity to the oxidation of the Fe at the substrate while the interior of the Fe resembles that of a single crystal, independent of the choice of substrate, being MAO or MgO. 

The measured magnetisation at remanence normalized to the saturation magnetisation $|M_{r}/M_{sat}|$ was calculated. While $|M_{r}/M_{sat}|$ was measured to be approximately 1 when probing along the Fe easy axis direction, $|M_{r}/M_{sat}|$ decreased to $1/\sqrt{2}$~=~0.707 when probing along the Fe hard axis direction. The $|M_{r}/M_{sat}|$ of all measured samples scatters on average below 10~\% around this value independent of substrate material, confirming a well defined magnetocrystalline anisotropy for Fe on both substrates. Thus, $|M_{r}/M_{sat}|$ for all the investigated Fe layers are well described as arising as a response from single crystalline Fe layers. 

\subsection{\label{sec:level3}Electronic transport properties}

\begin{figure}
\includegraphics{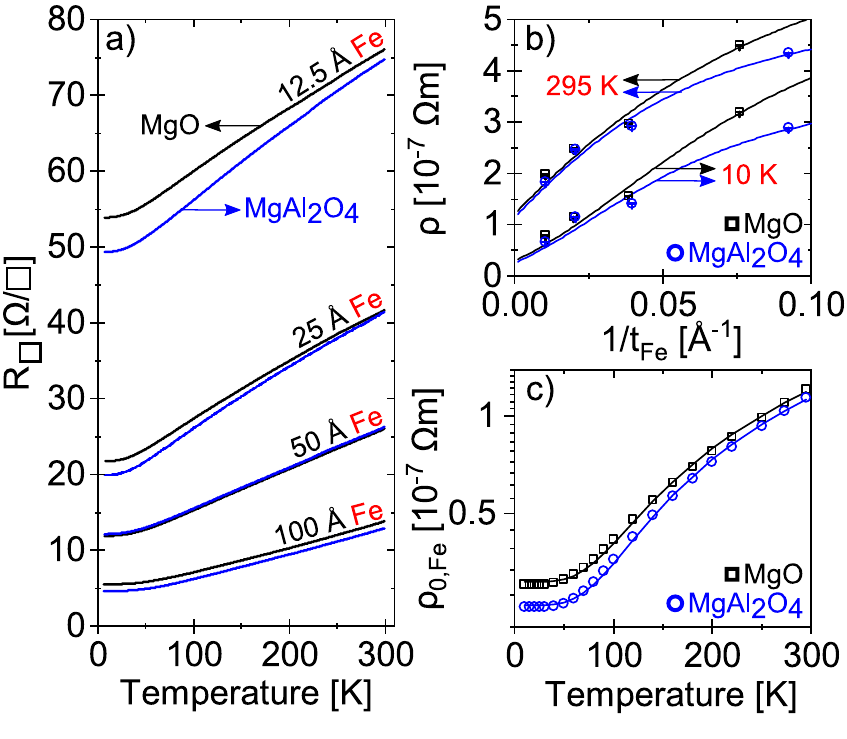}
\caption{\label{fig:transport} a) Temperature dependence of the sheet resistance $R_{\square}$ for MAO/Fe/Pd (blue) and MgO/Fe/Pd (black) bilayer films and for various Fe layer thicknesses. b) Representative resistivity values $\rho$ for the bilayers, for the temperatures of 10 and 295~K plotted against the inverse Fe layer thickness 1/$t_{\mathrm{Fe}}$. The lines are fits produced by the parallel resistor model described in the main text. Literature values for the points corresponding to 1/$t_{\mathrm{Fe}}$ = 0 are included from White and Woods \cite{Bulk_Fe_resistivities}. c) Fitting all resistivity data for the whole measured temperature range yields resistivities of the bulk with the same defect density as the Fe layers grown here, $\rho_{0, \mathrm{Fe}}$. The lines are fits resulting from the Bloch-Gr\"{u}neisen model, as described in the main text.}
\end{figure}

Fig.~\ref{fig:transport}a summarizes the determined sheet resistance of the films. It is important to keep in mind that the measurements were conducted on the whole stack of substrate/Fe/Pd, whereby the Fe thickness is varied while the Pd thickness of 50~{\AA} is kept constant for all samples. Fe and Pd have similar room temperature resistivities \cite{LandoltBornstein1983Fe, LandoltBornstein1983Pd}. The measurements show a typical metallic behavior of the samples with a reduction of resistance with decreasing temperature. The expected decrease in resistance with increasing film thickness is observed for both substrates. In addition to the total thickness of the stack increasing by increasing the Fe layer thickness, the sheet resistance of the stack is expected to decrease due to the smaller scattering contributions from the interfaces and surface.

In order to investigate the influence of the Fe layer crystal structure on the observed sheet resistance values of the Fe/Pd bilayers, we have employed a simple parallel resistor model \cite{Parallel-resistor}, representing the resistances of the Fe and Pd layers. As such, the resistivity $\rho$ of the bilayer is given by:

\begin{multline}
\begin{aligned}
        \rho = t\cdot R_{\square} &= \frac{t_{\mathrm{Fe}}+t_{\mathrm{Pd}}}{\left(\frac{1}{R_{\square,\mathrm{Fe}}} + \frac{1}{R_{\square,\mathrm{Pd}}}\right)} \\ &=(t_{\mathrm{Fe}}+t_{\mathrm{Pd}})\cdot \frac{\rho_{\mathrm{Fe}}\rho_{\mathrm{Pd}}}{t_{\mathrm{Fe}}\rho_{\mathrm{Pd}}+t_{\mathrm{Pd}}\rho_{\mathrm{Fe}}}
\end{aligned}
\end{multline}
where $R_{\square,\mathrm{Fe}}$, $t_{\mathrm{Fe}}$, $\rho_{\mathrm{Fe}}$ and $R_{\square,\mathrm{Pd}}$, $t_{\mathrm{Pd}}$, $\rho_{\mathrm{Pd}}$ are the sheet resistances, layer thicknesses, and resistivities for the Fe and Pd layers, respectively. In order to include contributions to the resistivities arising from the finite thickness of the layers, we express the Fe layer resistivities $\rho_{\mathrm{Fe}}$, as:

\begin{equation}
    \rho_{\mathrm{Fe}} = \rho_{0,\mathrm{Fe}}\left(1+K\frac{1}{t_{\mathrm{Fe}}}\right)
\end{equation}
with $\rho_{0,\mathrm{Fe}}$ the resistivity of the bulk with the same defect density as the Fe layer and $K=K'l_0$ a constant relating to the mean free electron path $l_0$ \cite{WEDLER19801}. To constrain the model fits with respect to bulk resistivity values, we included corresponding experimental data from White and Woods \cite{Bulk_Fe_resistivities}.

Fig. \ref{fig:transport}b, shows selected resistivity data for Fe/Pd layers on MgO and MAO for two temperatures, along with the resulting fitted curves. Analyzing the data over the entire measured temperature range and Fe layer thicknesses, results in an estimation of the temperature dependence of $\rho_{0,\mathrm{Fe}}$ for Fe on the two substrates, shown in Fig. \ref{fig:transport}c. A clear decrease of the values for the MAO substrates can be seen, correlating closely to the overall improved crystal structure presented in previous sections.

Finally, we test the validity of the Bloch-Gr\"{u}neisen model \cite{Ziman2001} on the extracted resistivity values for the Fe, shown in Fig. \ref{fig:transport}c, in the form of:

\begin{equation}
    \rho_{0,\mathrm{Fe}} = \rho_0 + A \left ( \frac{T}{\Theta_{\mathrm{D}}} \right )^5 \int^{\frac{\Theta_{\mathrm{D}}}{T}}_0 \frac{x^5}{(e^x-1)(1-e^{-x})}dx,
\end{equation}
where $\rho_0$ is the residual resistivity, $T$ temperature, $A$ a prefactor, and $\Theta_{\mathrm{D}}$ the Debye temperature \cite{Chowdhury_2013}. The fits for this formula are included in Fig. \ref{fig:transport}c and yield the values of 563 and 552~K for $\Theta_{\mathrm{D}}$, in the case MAO and MgO, respectively. In the literature the reported value for $\Theta_{\mathrm{D}}$ in bulk iron is 477~K, as derived from elastic constants data \cite{Stewart_1983}. Since $\Theta_{\mathrm{D}}$ is proportional to the interatomic force constants in the crystal structure of the material, the aforementioned values should be considered in the framework of the crystal structure and layer finiteness presented here. Both can imply significant changes in the elastic properties of the studied Fe films.


\section{\label{sec:level1}Summary and conclusions}

We find that thin Fe films with thicknesses between 12.5~{\AA} and 100~{\AA} can be grown with a well defined layering with a high degree of flatness. The Fe grown on MAO substrates has a significantly higher crystal quality which was shown by the observation of Fe Laue oscillations over a broad 2$\theta$ range in XRD as well as an about 40 times smaller mosaicity compared to Fe grown on MgO substrates. The Laue oscillations were observed for Fe on MAO with Fe layer thicknesses of and above 25~{\AA}. The thickest Fe layer of 100~{\AA} grown on MAO exhibits a bulk-like elasticity, which was confirmed by calculations of the Fe Poisson's ratio from the measured in- and out-of-plane Fe lattice parameters.

The obtained magnetic properties are only weakly reflecting the differences of crystal quality of the Fe layers. The saturation field along the Fe hard axis is lower for Fe grown on MAO compared to MgO, which is argued to arise from the tetragonal distortion of the Fe layers grown on MgO. The electronic transport in the Fe/Pd films of different thickness is dominated by finite size effects, which also show a phonon contribution.

The difference in crystal structure of Fe on MAO compared to MgO results in a difference in the electronic transport properties in the measured temperature range, with increased divergence at temperatures below 50~K and for Fe layer thicknesses of and below 25~{\AA}.
The difference in the bilayer residual resistance is not only attributed to the crystal structures of the Fe layers, but needs to account also for the difference in structure of the Pd capping layer on top. This is of importance for multilayered structures of technological interest as in the case of THz emission, where up to now studies had mainly focused on the Fe layer crystal quality \cite{Nenno_2019ip}.

Finally, the detailed characterization of the Fe layers presented here provides insights into the quality engineering of Fe/oxide superlattices \cite{Raanaei2008}, exhibiting intriguing properties with respect to interlayer coupling and switching \cite{Katayama2006, Moubah2016, Magnus2018}, as well as transport properties and \cite{Parkin2004, Yuasa_2004ig} magnetic domain control using electrical currents \cite{AuFeMgO_PhysRevRes_2021}.


\section*{Acknowledgments}
The authors would like to acknowledge financial support from the Swedish Research Council (Project No. 2019-03581 and 2019-05379). G.K.P also acknowledges funding from the Swedish Energy Agency grant 2020-005212.


\section*{Author Declarations}

\subsection*{Data availability}
The data that support the findings are available from the corresponding authors upon reasonable request.

\subsection*{Conflict of Interest}
The authors have no conflicts to disclose.


\bibliographystyle{elsarticle-num} 
\providecommand{\noopsort}[1]{}\providecommand{\singleletter}[1]{#1}%

\end{document}